# ITS and Real Time Cross Border Logistic Operations Optimization[1]

## Laura Coconea[1*], Miha Cimperman[2], Tobias Jacobs[3]

1. Swarco Mizar, Italy, laura.coconea@swarco.com
2. Institut "Jožef Stefan", Slovenija, miha.cimperman@ijs.si
3. NEC Laboratories Europe, Germany, tobias.jacobs@neclab.eu

**Abstract**

Moving parcels from origin to destination should not require a lot of re-planning. However, the vast number of shipments and destinations, which need to be re-aligned in real-time due to various external factors makes the delivery process a complex issue to tackle. Anticipating the impact of external factors though can provide more robust logistic plans which are resilient to changes.

The work described in this paper, was carried out in the EU-funded COG-LO project and addresses the issue of parcel delivery across the road network making use of context-awareness information as an input for the optimization operations. A positive impact derived from the implementation of these services is expected due to complex event detection, context awareness and decision support at both local and global level of logistics operations.

**Introduction**

In logistics, parcel management is strongly connected to the transportation environment: vehicles are routed for parcel pick-up and delivery in the optimal manner across the road network. Customers expect fast and reliable services with reasonable prices. As a result, there is a need in having the optimal plan and to execute the routing efficiently and therefore various challenges are present.

Once planned, functionally moving a parcel from origin to destination should not require a lot of re-planning. However, the complexity arises from the vast number of shipments and destinations, which need to be re-aligned in real-time due to various external factors. This is particularly difficult in a deterministic lead-time and the finite amount of available resources. Consequently, the optimization of properties like delivery time, resource utilization and geographical coverage is an inherent challenge of large-scale logistics operations.

This work, developed in the EU-funded COG-LO (1) project, tackles the issue of parcel delivery across the road network making use of context-awareness information as an input for the optimization operations. The document starts by introducing the motivation of such a work and continues with a second section where after a brief overview of the COG-LO project, continues with the description of various situational awareness services (with a focus on Traffic Management) along with the description of how these services enable the optimization of parcel delivery. Last but not least, the paper documents the expected impacts of this approach while conclusions and next steps close the document.

---

[1] This work has been included in the proceedings of the Virtual ITS European Congress, 2020.

**Motivation**

Once planned, functionally moving a parcel from origin to destination should not require a lot of re-planning. However, the complexity arises from the vast number of shipments and destinations, which need to be re-aligned in real-time due to various external factors. This is particularly difficult in a deterministic lead-time and the finite amount of available resources. Consequently, the optimization of properties like delivery time, resource utilization and geographical coverage is an inherent challenge of large-scale logistics operations.

A number of real-time scheduling algorithms has been developed by the logistics community since the 1980's; they were essentially Decision Support Systems (DSS) and included typical DSS elements. During the 1990s, GIS emergence permitted the display and manipulation of spatial information, and, thus, supported the realization of more comprehensive models of the road network, allowing more realistic modelling of path constraints. Dynamic real-life problems often require rich models; however, in most of the literature on dynamic routing problems, like the dynamic full-truckload pickup and delivery problem, some simplifying assumptions are made for relaxing the stochastic nature of those problems. Currently, most of the dynamic routing algorithms lack or ignore the random nature of external factors, which affect the delivery efficiency, and rely only upon online information updates for periodic re-optimization. In fact, periodic re-optimization has become extremely popular and is a reactive approach that tends to correct the operational planning when changes have already happened.

In periodic re-optimization (or rolling-horizon optimization), dynamic routing reacts to the observations from the real-world operations and does not anticipate proactively the implications from external factors such as (i) missing a delivery because of the recipient's absence or (ii) delivery demand fluctuations during the day. Anticipating the impact of external factors though can provide more robust logistic plans which are resilient to changes. Hence, it is desirable to model the dynamic behaviour of demand variation, capacity and operational capability of logistics operators. This way, demand fluctuations and missed deliveries can be treated more efficiently by enabling logistic operators to create more reliable plans that require less re-alignments and anticipate the operational variations. Apart from reliable planning, real-time re-alignment of the plans to the existing conditions is needed.

**ITS and Real Time Cross Border Logistic Operations Optimization**

**Tackling the challenge: the COG-LO approach**

*Introduction to COG-LO*

The main goal of COG-LO is to create the framework and tools that will add cognition and collaboration features to future logistics processes by: 1) Introducing the Cognitive Logistics Object (CLO) by adding cognitive behavior to all involved Logistics actors and processes; 2) Developing the environment that will allow CLOs to exchange information through ad-hoc social secure networks. The main objectives of COG-LO are: (a) Definition of a Future Cognitive Logistics Objects reference implementation model that supports cognitive logistics behaviour; (b) Design and develop the necessary analytics and cognitive tools that enable complex event detection, context awareness and multi-criteria decision support. (c) To design and develop the collaboration platform based on hybrid ad-hoc social networks of CLOs. (d) To design and develop the COG-LO tools (Cognitive Advisor, which realizes the cognitive behavior of CLO based on the reference implementation model, and Tweeting CLOs, allowing CLOs to exchange information in ad-hoc Social IoT networks); (e) To evaluate the COG-LO results via intermodal, cross-country and urban logistics pilot operations. (f) To introduce new business models for ad-hoc collaborations affecting all logistics stakeholders and create a community of Logistics, ICT and urban environment stakeholders. The consortium is led by SingularLogic and consists of 14 partners from 8 countries.

COG-LO aims to introduce a breakthrough approach in terms of transport execution, based on (i) the new cargo-centric multi-modal transport models (ii) a decision criteria function model which will provide the CLOs multi-criteria decision-making capability based on different criteria like SLA agreements, current transport status, delivery destination and time; (iii) complex event detection, context awareness and decision support at both local and global level and (iv) the exploitation of ad-hoc social networks, for a seamless freight transport execution.

*Context Awareness*

The potential value-added services provided by context awareness tools, which enable real time information exchange and intelligent interaction among stakeholders that make use of smart infrastructure (who manages the transport infrastructure and logistics, who transports the goods, who charges/discharges and who controls it), are the following:

- providing information of transport times, estimated times of arrival and anomalous traffic, queues and accidents (in order to enable the best decisions for the planning and execution of travel);

- exchanging information and data between logistics operators in favour of greater integration;

- reducing logistics downtime, facilitating the transport workflow, managing access control, document exchange and clearing operations;

**ITS and Real Time Cross Border Logistic Operations Optimization**

- detecting and predicting critical events that effect road infrastructure delivery, such as traffic congestions, increased travel time due to external (weather) conditions and changed physical constraints for travelling (reduced speed limit, etc.).

- estimating effect size (size of infrastructure effected) on event, yielding a number of objects to be involved in logistics optimization

- facilitating the planning and meeting of supply and demand with emphasis on transport and intermodal forms of transport.

Smart infrastructure can be divided into two parts: digital and physical. It is important to distinguish between these two, as they are so different in characteristics. Physical infrastructure is not only the roads vehicles travel on. It consists of roads, tunnels, ports, warehouses, terminals and similar asset-based facilities. Toward this physical smart infrastructure, smart vehicles can have two-way information flow through the ICT systems that are part of the digital infrastructure. In this way the vehicle is always connected, providing real-time information as needed. The smart digital infrastructure retrieves, manipulates, stores and communicates data and information from the physical infrastructure to and from the smart vehicles using different digital technologies such as sensors, cameras, databases, and positioning technologies. The smart infrastructure enables peer-to-peer information exchange about the goods, vehicles and infrastructure to be communicated between participants as needed. At the same time, the emerging connectivity and anticipated automation is expected to revolutionize the future traffic management and control systems. The customers' expectations are increasing hand in hand with the unprecedented data availability and technological developments. The new data sources (e.g. connected vehicles, cameras) and new technologies (IoT, Big Data, AI) open an opportunity for a fully collaborative traffic management supporting the supply chain.

A positive impact derived from the implementation of these services is expected due to complex event detection, context awareness and decision support at both local and global level of logistics operations. The approach is supported by the deployed technological framework with a unified solution for accessing real-time information, under a common semantic framework for interoperable data integration, which will lead to seamless freight transport execution.

Based on the framework for dynamic planning and optimization of logistics operations, the project will offer tools for optimization of logistics operations in terms of delays (time to deliver), cost per mile or parcel delivered and increasing distribution vehicles utilization (increasing load factor). It will also enable service operators to offer higher service quality to their customers, due to reduced transit times and improved efficiency. The listed aspects will enable increased Quality of Services.

Within COG-LO, the output of the Traffic Management services presented in this document, represents an input for the Cognitive Advisor Tool (CA), for infrastructure abstraction optimization services of logistics operations.

The context awareness services developed in COG-LO aim to tackle the main challenges, by

**ITS and Real Time Cross Border Logistic Operations Optimization**

providing information of decision making in logistics optimization for CA orchestration. ITS applied to logistics is a topic supported by the European Commission, that for example in its "Freight Transport Logistics Action Plan" (2) encourages the use of corridors "to experiment environmentally-friendly, innovative transport units, and advanced ITS applications." Also, initiatives like COGISTICS (3), COG-LO, URSA MAJOR (4) support and deploy the same approach in optimizing logistics operations.

Following these guidelines, in COG-LO, the role of the Traffic Management System is to provide Real Time Traffic Information (RTTI) related to the road situation and to the traffic management measures, in order to enable the optimization of the delivery routes and in order to achieve convergence between the Traffic Management strategies and the individual needs.

The features made available by the TMS (Traffic Management System) are divided in four categories:

1. Traffic Events Management

- Automatic event detection (LoS)
- Integration of external event sources (DATEX)
- Manual event configuration/ management
- Automatic response plan generation
- Driver awareness
- Publication of Real Time information

2. Travel Times Management

- Calculation of Travel Times
- Identification of LoS
- Publication of Real Time information

3. Traffic Management Strategies

- Situation monitoring
- Strategy definition
- Strategy control
- Manual control

4. V2X: In Vehicle Information input to CLOs/Drivers

- Traffic Lights information available real time

**ITS and Real Time Cross Border Logistic Operations Optimization**

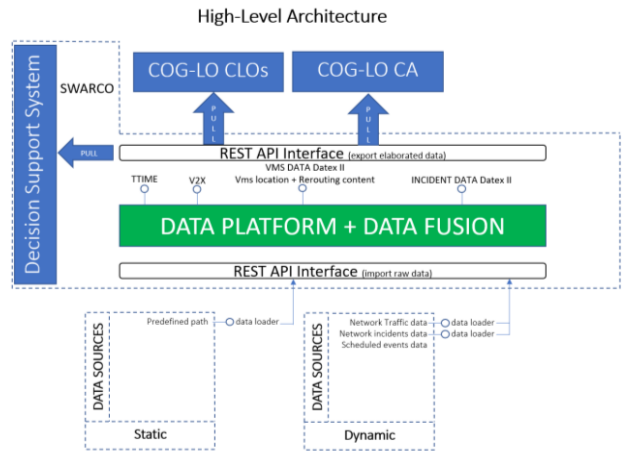

Figure 1 Functional architecture of COG-LO TMS

*Optimization*

Optimization services play a key role in the COG-LO infrastructure, enabling the Cognitive Logistic Objects (CLO) to make informed decisions and proposals in reaction to events that occur in the physical world and trigger virtual events.

Optimization Services are components in the COG-LO system that have the ability to solve optimization problems optimally or near-optimally and propose solutions. While each use case has unique variants of optimization problems to solved, both real-time and in batch mode, the COG-LO project has performed a unified problem analysis across several pilots to identify the set of needed optimization modules and their capabilities. The most obvious commonality among all use cases is that each one contains some form of the Capacitated Vehicle Routing Problem (CVRP). In the CVRP, we are given a set of vehicles with limited capacities and a set of orders, where each order is associated with a size and a node in a graph. Packing problems describe the task of assigning a set of items, each having a certain size, into a number of containers each having a maximum capacity. In assignment problems, each element from a set A has to be assigned to a set B, so as to minimize a cost function.

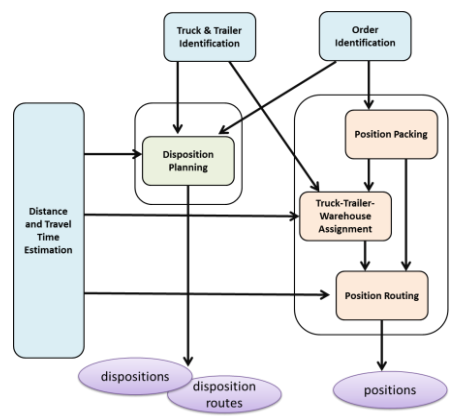

Figure 2: Optimization Service Architecture for Multi-Modal Logistics

**ITS and Real Time Cross Border Logistic Operations Optimization**

An example of the Interplay of Optimization Services is shown in Figure 2, which depicts the optimization tasks of a Multi-Modal logistics use case, where the process to be optimized consists of Dispositioning (collecting parcels at customer sites) and Positioning (consolidation into larger transport units, and determining multi-modal routes for these units).

COG-LO is developing robust and adaptive optimization solutions, coupling analytics and optimization, and considering the effect of optimization control to the performance of operations in an environment with continuous external variations. Using historical data about missed deliveries, daily delivery fluctuations, the logistic plans anticipate operational noise due to external factors.

*Cognitive Advisor (CA)*

The Cognitive Advisor tool (CA) represents an agent for handling events in logistics infrastructure and generates interventions in the form of new distribution routes for vehicle on the field. The Cognitive Advisor also consolidates all analytical components and knowledge services developed by the COG-LO project and adds a concept of event orchestration, and infrastructure abstraction, to form a distributed modular structure of logistics intelligence. The CA operates on data describing physical infrastructure, parcel data, events data, and business data. Moreover, the data consumed by the CA has been structured into two main categories, namely: infrastructure data and external data (traffic infrastructure and event data).

CA primarily builds a graph representation of logistics infrastructure, based on street infrastructure, post office locations, events, physical limitations and other constraints that effect optimization of logistics. Together it forms a layer for constructing digital representation of infrastructure and optimization layer of logistic processes. CA is utilized either on initial daily fleet orchestration, or in case of new ad-hoc event (real time optimization), such as new pickup/delivery request, traffic event, vehicle break down, etc. The main service exposed therefore enables creating new Route recommendations in case of new events in COG-LO infrastructure. It turns out that the set of optimization problems encompasses several variants of the capacitated vehicle routing problem, variants of assignment problems, and packing problems. Based on use case analysis, the functional and non-functional requirements for the optimization Module have been derived. While CA's architecture is designed modular, new use cases can be added or upgraded with either optimization functionalities, new instantiation of optimization modules or adjusting the process flow.

The CA is designed as a modular structure of decision making, being able to process ad-hoc events, as well as daily requirements of parcel distribution and estimate most optimal distribution schema for logistics intraday operations.

**ITS and Real Time Cross Border Logistic Operations Optimization**

**Expected impact**

By enabling ad-hoc event processing, increasing response time and dynamic parcel delivery optimization, the assisted parcel routing should enable several beneficial impacts on key performance indicators, such as:

- Improving Load factor – we can expect better load factor results since there will be an ongoing process of shipment consolidation. One portion of cross border (XB) shipments will be transferred directly through the border which will result in a decrease of shipment volumes transferred via the existing routes of shipment exchange.

- Decreasing total route length – since the current routes between the offices of exchange are fixed and not subject to any change we shall see a significant decrease in route length due to the CA's employment of the existing infrastructure and fleet near the border.

- Lower fuel consumption – fuel consumption will be lower since all of the shipments near the border will be transferred directly to the post offices on the other side of the border. With the newly established XB postal-logistic chain we will bypass existing routes which go through the logistics center and the office of exchange.

- Decreasing total costs – Bypassing of the current route of shipment exchange will result in shorter transfer times and shorter routes between the involved actors. We will have lower costs maintaining the XB postal chain between the operators

- Better customer satisfaction/customer experience – A major impact is expected on the internal management of infrastructure, fleet portfolio and other resources. This installment of the XB postal-logistics chain will benefit both postal operators with the opportunity to offer new services. New services will have a substantial effect on the end-users.

**Conclusions**

The work described in this paper, was carried out in the EU-funded COG-LO project and addresses the issue of parcel delivery across the road network making use of context-awareness information as an input for the optimization operations.

In the general parcel workflow management, the proposed solution will contribute by optimizing order management by means of assigning the right order to the right vehicle and deciding whether this vehicle needs to be re-routed in order to meet the ETAs. Indeed, in the general case, after the courier has shipped the scheduled orders for delivery from the sorting centre and while being on route, a new order may be received. In such occasion, the courier should accept the new ad-hoc order and possibly re-arrange the deliveries and change the scheduled route without affecting the daily orders logistics. The system will also respond to the dynamics of operational constraints and adjust the parcel flow with potential new routes recommendations.

**Acknowledgements**

The COG-LO project (Cognitive Logistics Operations through secure, dynamic and ad-hoc collaborative networks) has received funding from the European Union's Horizon 2020 - EU.3.4. SOCIETAL CHALLENGES - Smart, Green and Integrated Transport programme under grant agreement number 769141.